\documentstyle[aps,prl,graphicx]{revtex}

\newcommand{\be}{\begin{eqnarray}}
\newcommand{\en}{\end{eqnarray}}

\begin{document}
\draft
\twocolumn[\hsize\textwidth\columnwidth\hsize\csname 
@twocolumnfalse\endcsname
\title{Staggered flux fluctuations and the quasiparticle scattering rate
in the SU(2) gauge theory of the $t$-$J$ model}
\author{Carsten Honerkamp and Patrick A. Lee}
\address{Department of Physics, Massachusetts Institute of Technology, Cambridge MA 02139, USA } 
\date{December 4, 2002}
\maketitle
\begin{abstract} 
We study staggered flux fluctuations around the superconducting state of the SU(2) meanfield theory for the two-dimensional $t$-$J$-model and their effect on the electron spectral function. The quasiparticle peaks near $(\pi,0)$, $(0,\pi)$ get strongly broadened and partially wiped out by these fluctuations while the quasiparticle peaks near the nodes of the $d$-wave gap are preserved over a wide parameter range. The strength of these effects is governed by an energy scale that decreases towards zero for doping $x \to 0$ and that is related to the energy splitting between the SU(2)-related superconducting and staggered flux meanfield states.
\end{abstract}
\pacs{}
\vskip1pc]
\narrowtext

By now several basic properties of the high temperature superconducting cuprates are well established. From the Mott insulating antiferromagnet to the highly overdoped state there exists a large number of experiments some of which can be understood within a particular framework -- like Heisenberg-type spin models for the undoped compounds\cite{manousakis} or $d$-wave BCS theory for the superconducting state at low temperatures\cite{wang} -- and some others that remain subject of lively debates, such as the Nernst effect above the superconducting transition temperature\cite{ong} or the $d$-wave-like pseudogap seen in angular resolved photoemission\cite{damascelli}. The intriguing question that is raised by the latter experiments is the connection and relation between the different phases of the high-$T_c$ cuprates. 
A theoretical framework that incorporates and partially predicted most of the observed phases is the gauge theory approach to the doped two-dimensional (2D) $t$-$J$ model\cite{lee-review}.
The mean-field theory of this description includes a spin- or pseudogap phase at temperatures above a $d$-wave superconducting state and renders a superfluid weight and critical temperature scaling with the doping $x$, i.e. the connection to the Mott insulating state is described in a natural way.

Apart from these successes of the meanfield theory, the 2D $t$-$J$-model gives rise to even richer physics. This became clear in the search for a variational ground state at zero doping: at $x=0$ there is a huge degeneracy of spinon wave functions corresponding to different meanfield decouplings and being related to each other by SU(2) gauge transformations\cite{affleck}. After projecting onto the physical Hilbert space with one particle per site, all these states are in fact equivalent. Of course in the doped system this degeneracy is broken, leaving the superconducting state as the ground state. Nevertheless the energy splitting between these states represents a new small energy scale in the underdoped system. 
This idea led Wen and Lee\cite{wenlee} to formulate the SU(2)-gauge theory for the doped $t$-$J$ model. This theory attempts to incorporate the fluctuations between the different mean-field states also for nonzero $x$. Recently Lee and Nagaosa\cite{collmodes} analyzed Gaussian fluctuations around the superconducting state of the SU(2) meanfield theory. They found that next to the conventional collective modes of a $d$-wave superconductor there are other excitations which correspond to the gauge degrees of freedom, i.e. to fluctuations between the superconducting state and other meanfield states that would be degenerate at $x=0$ or above the boson condensation temperature $\sim xt$. 
The largest spectral weight at low energies of these new collective modes is in the so-called $\theta$-mode that corresponds to fluctuations into the staggered flux (SF) state. The static SF state at finite doping\cite{hsu} breaks the time-reversal and translational symmetry as the hopping matrix element acquires a staggered imaginary component. This generates alternating currents $\propto x$ around the plaquettes of the square lattice. 
The quasiparticle spectrum of the SF state is similar to that of the $d$-wave superconductor with the difference that the SF state forms a small Fermi surface at finite doping. 
It is has been suggested\cite{kishine} that vortices in the underdoped system may support SF cores, as the energy of the static SF state is only slightly higher than the superconducting state that is locally destroyed by the penetrating magnetic field, making the SF vortex energetically  cheap. By the same argument the proliferation of vortex/anti-vortex-pairs with SF cores may represent an effective way to destroy superconductivity with increasing temperature\cite{lee-orbicur}. The existence of sizable SF correlations in the lightly doped $t$-$J$ model has been shown by an analysis of Gutzwiller projected $d$-wave superconducting wavefunctions\cite{ivanov} and exact diagonalization\cite{leung}.

In this work we analyze the coupling of Gaussian SF fluctuations to the quasiparticles in the superconducting state. With increasing temperature we find a strong anisotropic degradation of the quasiparticles around $(\pi,0)$ and $(0,\pi)$, the stronger the smaller the hole doping is. The quasiparticles in the nodal directions remains sharp up to much higher temperatures. This trend is a direct consequence of the SU(2) symmetry at zero doping and in qualitative agreement with the results of photoemission experiments\cite{valla}. Furthermore it is consistent with the idea that the pseudogap state actually fluctuates between all these states and the superconductivity develops out of the pseudogap state by freezing out these fluctuations.

The SU(2) gauge theory for the 2D $t$-$J$ model\cite{wenlee,kishine} enlarges the original Hilbert space by writing the local electron operators in terms of two spin-$\frac{1}{2}$ fermions $f_{i,\uparrow}$, $f_{i,\downarrow}$ and two charged bosons $b_{i1}$ and $b_{i2}$. 
These degrees of freedom are coupled through the temporal components of SU(2) gauge fields, $a_{0,\ell}$, $\ell = 1,\dots,3$. Integration over these gauge fields is equivalent to projecting onto the physical Hilbert state.
In meanfield theory for the underdoped state the superconducting transition occurs due to condensation of the bosons at temperatures $T \sim xt$. In the following we will focus on temperatures below this transition and assume fully condensed bosons. We choose the radial gauge, where only the $b_{1,i}$ boson acquires a nonzero expectation value and $\langle b_{2,i} \rangle =0$. 
The meanfield solution for the superconducting state is characterized by nonzero expectation values for fermion hopping, $\chi_{ij}$, fermion pairing, $\Delta_{ij}$, and the local boson condensate $b_0= \sqrt{x}$.
The Lagrangian can be written as\cite{collmodes}
\begin{eqnarray}
L &=& {\tilde{J} \over 2} \sum_{<ij>} Tr \left[ U_{ij}^\dagger
U_{ij} \right] +
\tilde{J}  \sum_{<ij>} \left( \Psi^\dagger_{i}
U_{ij} \Psi_{j} +
c.c. \right) \nonumber \\
&+&  \sum_{i\sigma} f^\dagger_{i,\sigma} \left(
\partial_\tau - a_{0i}^\ell \tau^\ell \right) f_{i,\sigma} 
-{x\over 2}\sum_{<ij>,\sigma} t_{ij} 
f^\dagger_{i\sigma} f_{j\sigma}
\,\,\, . \label{Lagrange} \end{eqnarray}
The $\Psi_{i,\sigma}$ denote the fermion doublet $(f_{i,\uparrow}, f^\dagger_{i,\downarrow})$.  In the meanfield approximation
the $a_{0,i}^\ell$ are chosen to fulfill the constraint of no double occupancy on average. 
With our gauge choice only $a_{0,3}$ deviates from zero. 
$\tau^\ell $ are the Pauli matrices.
The matrices $U_{ij}$ contain the meanfields on the nearest neighbor bonds 
$\langle ij \rangle$ that arise from decoupling the exchange term, $\tilde J = \frac{3}{8} J$. In the radial gauge the superconducting meanfield solution is given by
\be
U_{ij}=
\pmatrix{-\chi_{ij}^\ast&\Delta_{ij}\cr
\Delta_{ij}^\ast&\chi_{ij}} 
\en
with $\chi_{ij}=\chi_0$ and $\Delta_{ij}=\Delta_0 (-1)^{i_y+j_y}$.
At $x=0$ this solution is degenerate with the SF solution that has a staggered imaginary hopping amplitude $W_{ij} \propto i \theta (\vec{x}_i) (-1)^{i_x+i_y} (-1)^{i_y+j_y}$ replacing the pairing amplitude $\Delta_{ij}$. Here $(-1)^{i_x+i_y}$ takes care of the staggering, $(-1)^{i_y+j_y}$ is the $d$-wave sign.

\begin{figure}
\includegraphics[width=.49\textwidth]{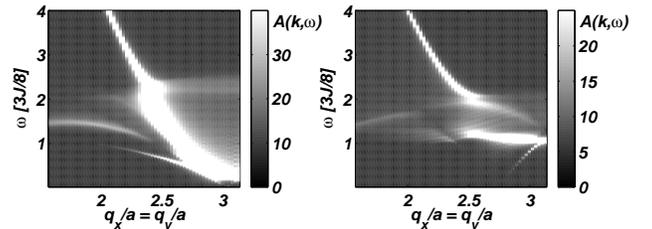}
\caption{The spectral function of the staggered flux fluctuations for 2$\%$ doping (left plot) and $10\%$ doping (right plot) along the line $q_x=q_y$. Further parameters $J=0.25t$ and $t'=-0.45t$. $\omega$ is measured in units of $\tilde J =\frac{3}{8}J $}
\label{specplot}
\end{figure}

The SF fluctuations correspond to a adding a fluctuating staggered imaginary part to the diagonal $U_{ij}$-matrix elements, i.e. a perturbation of the type $\delta U_{ij} = i \theta (\vec{x}_i) (-1)^{i_x+j_y} \tau^0$, where $\tau^0$ denotes the identity matrix. Combining Matsubara frequencies and wavevectors into one index $q=(i\nu, \vec{q})$, the $\theta$-dependent part of the action reads
\be
S_\theta = T\, \sum_q \theta_q \Pi_q \theta_{-q} + \frac{T}{N} \sum_{k,q} \,  g(k,q)\, \Psi^\dagger_{k+q} \tau_0 \Psi_k \, \theta_q   \, , \label{thetaaction}
\en
where $N$ denotes the number of lattice sites and the nearest neighbor formfactor is ($\vec{p}=\vec{k}+\vec{q}/2$, $\vec{Q}= (\pi,\pi)$)
\be g(\vec{k},\vec{q}+\vec{Q} ) = 2i \left[  e^{-iq_x/2}\cos p_x  -e^{-iq_y/2} \cos p_y \right] \, . \label{formfactor} \en
The kernel $\Pi(\vec{q},\nu)$ can be found from integrating out the fermions as described in Ref. \onlinecite{collmodes}. The spectral function obtained by inverting $\Pi(\vec{q},\nu)$ is shown in Fig. \ref{specplot}.
The essential points to notice are: {\em a)} there is a a relatively well defined massive mode with a gap of $\sim 10x\tilde J$; {\em b)} its low energy spectral weight is concentrated around the wavevector $(\pi,\pi)$, justifying the designation {\em staggered} flux mode as pars pro toto. Note that due to the SU(2) symmetry at $x=0$ the SF mode energy at $(\pi,\pi)$ comes down to zero energy for $x\to 0$. Therefore the impact of this mode on the quasiparticles increases drastically for $x \to 0$.  

In the following we will describe the effects of the SF fluctuations on the lifetime and the weight of the superconducting quasiparticles. We use perturbation theory in the effective interaction mediated by the SF fluctuations. In first order we have two diagrams, the Hartree and the Fock term. The latter contains the information on lifetime and wavefunction renormalization. The fermion selfenergy given by the Fock diagram is
\begin{equation} \Sigma (k) =  \frac{1}{4}\sum_{q} |g(\vec{k},\vec{q})|^2 \Pi^{-1} (q)
G_f (k+q) \, \label{self} \end{equation}
with the fermion Green's function in the superconducting state
$G_f (\vec{k},i \omega) = u_{\vec{k}}^2/[i \omega - E_{\vec{k}}] + v_{\vec{k}}^2/[i \omega + E_{\vec{k}}] $.
The fermion dispersion is $E_{\pm}(\vec{k}) = \pm \sqrt{\xi (\vec{k}) +\Delta (\vec{k})}$ with $\xi (\vec{k}) = -2xt( \cos k_x -\cos k_y) -4xt' \cos k_x \cos k_y $. 
By analytic continuation of the external fermion frequency we can then obtain real and imaginary parts of the fermion self energy. Im$\Sigma (\vec{k},E)$ is responsible for a finite lifetime of the quasiparticle with excitation energy $E$, while the quasiparticle weight $Z_{\vec{k}}$ is given by\cite{mahan}
$Z_{\vec{k}}^{-1} =  1-  \mbox{Re} [ \Sigma(\vec{k},E) - \Sigma(\vec{k},-E)] /2E $. 

In the SU(2) meanfield theory the unperturbed spectral function of the physical electron, $A(\vec{k},\omega)$, consists\cite{wenlee} of a coherent quasiparticle peak at energy $E_{\vec{k}}$ with weight $x$ and a weak incoherent background that sets in above $E_{\vec{k}}$. We include the imaginary part of the selfenergy and the wavefunction renormalization $Z_{\vec{k}}$. This broadens the quasiparticle peak by the inverse lifetime $Z_{\vec{k}}\, \mbox{Im} \Sigma (\vec{k},E)$ and decreases its weight to $Z_{\vec{k}}\, x$. We neglect the renormalization of the dispersion as this would require a selfconsistent treatment of the SF fluctuations and the meanfields on the bonds. Our approximation is appropriate if the renormalized quasiparticle dispersion is not too different from the the non-interacting one.
We do not include selfenergy effects on the incoherent background. The only effect that we may miss this way is the increase of the spectral weight in the incoherent spectrum through a transfer of weight when the coherent peak is diminished by a small $Z$-factor.

\begin{figure}
\begin{center} 
\includegraphics[width=.49\textwidth]{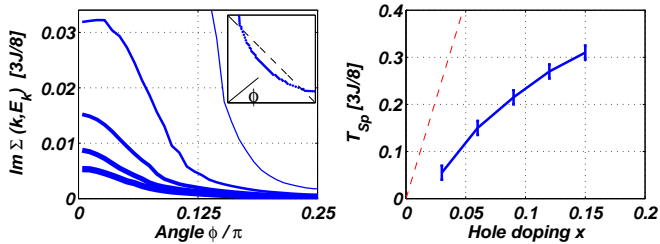}
\end{center} 
\caption{Left: Imaginary part of the selfenergy at the quasiparticle energy along a trajectory near the points of minimal excitation energy for angle $\phi$. The inset shows these points for $x=0.15$. The doping levels are $x=0.03$ (thinnest line), $0.06$, $0.09$, $0.12$, and $x=0.15$ (thickest line), $T=0.05\tilde{J}$. Right:  Temperature where the quasiparticle peak is suppressed down to the height of the incoherent background versus doping $x$. The dashed line shows the bose condensation temperature. }
\label{xang}

\end{figure}
\begin{figure}
\begin{center} 
\includegraphics[width=.49\textwidth]{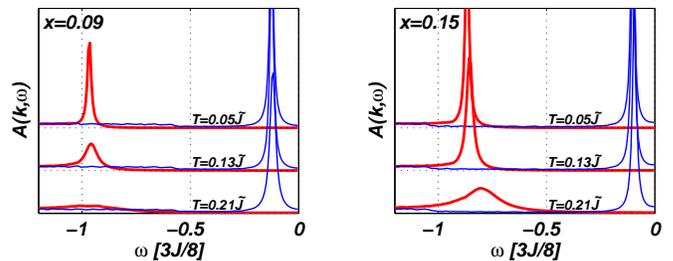}
\end{center} 
\caption{$A(\vec{k},\omega)$ in presence of the staggered flux fluctuations for dopings $x=0.09$ (left) and $x=0.15$ (right) and different temperatures. 
The thick (thin) lines corresponds to $\vec{k}$ close to $(\pi,0)$ (the nodes of the superconducting gap).}
\label{specfun}
\end{figure}

 
We have calculated the spectrum of the SF fluctuations for different dopings and temperatures. We choose $J=0.25t$ and $t'=-0.45t$. 
With increasing underdoping the SF mass gap gets smaller and the mode starts to show impact on the quasiparticle scattering rate. As shown in the left plot of Fig. \ref{xang} this first occurs for the high energy quasiparticles around  saddle points, for smaller doping the decay rates of quasiparticles closer to the nodes of the $d$-wave gap grow as well. The suppression of the quasiparticle weight is stronger at the saddle points as well, but the anisotropy is less pronounced. With increasing temperature the scattering phase space of the quasiparticles and the population of the finite frequency mode becomes larger. This causes the decay rates to grow considerably with increasing temperature, most strongly in the saddle point regions. 
 In Fig. \ref{specfun} we compare the quasiparticle peaks close top $(\pi,0)$ and $(0,\pi)$ and near the nodes of the $d$-wave gap  at dopings $x=0.09$ and $x=0.15$ for three different temperatures.
If we increase $T$ for $x=0.09$ the antinodal quasiparticle peaks near $(\pi,0)$ and $(0,\pi)$ are nearly wiped out and almost merge into the continuum. For $x=0.15$ and the same temperatures the peak is still clearly visible. 
In fact we expect the full incoherent background to be higher than that of the mean field theory due to the transfer of spectral weight from the coherent part and additional scattering processes. In the right panel of Fig. \ref{xang} we sketch the doping dependence of temperature above which the spectral peak of the antinodal quasiparticles gets smaller than the background together with the boson condensation temperature taken from Ref.\cite{wenlee}.

The strong anisotropy of the scattering rate can be understood as combination of two effects. First, for the main scattering vectors $\approx (\pi, \pi)$ the formfactor $(\ref{formfactor})$ of the coupling between SF fluctuations and fermions is basically $\propto \cos k_x -\cos k_y$. Thus the quasiparticles near the nodes of the $d$-wave gap are protected from the scattering to some degree. Second, the quasiparticles around $(\pi,0)$ and $(0,\pi)$ are high energy excitations and have therefore a much larger phase space to decay than the low energy excitations close to the nodes. 

The growing strength of the SF fluctuation with rising temperature or decreasing hole density will certainly affect the superconducting long range order. An effective path to destroy the phase coherence may be the proliferation of vortex/anti-vortex pairs with SF cores which are expected to be energetically cheap local excitations\cite{kishine,lee-orbicur}. In this case the fermion pair field $\Delta_{ij}$ will develop strong phase fluctuations and will be locally suppressed in exchange for a nonzero SF amplitude in the diagonal components of the $U_{ij}$-matrix. Our present analysis that considers on the SF fluctuations only does not allow a precise estimate of the superconducting transition temperature due to this mechanism. In any way it will approach zero with decreasing doping as the energy cost for a SF vortex disappears towards $x=0$.

For our perturbative approach to be valid there needs to be a nonvanishing temperature window where the meanfield picture makes sense.
As a consistency check, we estimate the effect of SF quantum fluctuations on the meanfield pairing amplitudes $\Delta_0$ and $\chi_0$. 
We calculated the free energy of the meanfield state plus the Gaussian SF fluctuations as a function of the size of the pairing and hopping meanfields. 
The free energy contribution of the SF fluctuations is lowered roughly linearly by decreasing $\Delta_0$, i.e. the fluctuations tend to reduce the pairing meanfield.  By extrapolating the linear $\Delta_0$-dependence of the fluctuation contribution and the quadratic variation of the meanfield energy, we can estimate the renormalized value for $\Delta_0$. At zero temperature and $x=0.06$ this methods indicates to a significant renormalization of $\Delta_0$ down to $55 \%$ of its meanfield value. For $x=0.1$ the renormalization is only down to $80\%$. The suppression of $\chi_0$ turned out to be much weaker. 
Although this is only  a rough estimate it tells us two things: it is likely that the superconducting ground state is stable against SF fluctuations for some nonzero $x$, and yet the SF fluctuations may be strong enough to cause visible effects. This is in good agreement with our selfenergy calculation.

The quasiparticle scattering by the SF fluctuations has some similarities with the magnetic mode picture\cite{norman} constructed around the spin resonance in the superconducting state. Like the spin mode we expect the SF fluctuations to be sharply defined only in the superconducting state. 
Note that in our case the spectral function of the SF fluctuations and their coupling to fermions follow directly from the formalism and no additional assumptions have to be made. The magnetic resonance is also present in the gauge theory picture\cite{brinckmann} and antiferromagnetic ordering for $x\to 0$ can be understood as instability of the fluctuating SF state\cite{dhkim}. 
Possible distinctions and combination effects of these two types of fluctuations deserve further consideration.  

In conclusion we have shown that the SU(2) meanfield description of the 2D $t$-$J$ model suggests an important role of staggered flux fluctuations in the superconducting state. The SF fluctuations cause a strongly anisotropic degradation of the superconducting quasiparticles, consistent with angle-resolved photoemission\cite{valla}. With decreasing doping the quasiparticles in the saddle point regions near $(\pi,0)$ and $(\pi,0)$ get increasingly destroyed, while the quasiparticles in the nodal directions near $(\pi/2,\pi/2)$ remain relatively well defined. A related tendency is observed at fixed doping when the temperature is increased: the quasiparticles at the saddle points get broadened quickly, while the spectral functions around the nodes of the superconducting gap remain sharp over a much larger temperature range. These trends are consistent with the doping dependence of the energy separation between superconducting and SF meanfield states that vanishes for $x \to 0$. We stress that this energy scale is a direct consequence of the local moment formation giving rise to the local SU(2) symmetry\cite{affleck} in the $t$-$J$ model at zero doping. 

If we assume that the superconducting transition goes along with a rapid increase of SF fluctuations (combined with superconducting phase fluctuations, \cite{lee-orbicur}), this opens a new way to understand the spectral functions in the pseudogap regime above the superconducting transition. Then, near $(\pi,0)$ the quasiparticle peaks will be wiped out, while along the Brillouin zone diagonal remnant peaks may survive. Our perturbative calculation around the meanfield state is too simple to describe the superconducting transition. Nevertheless it shows that strong SF fluctuations can produce an anisotropic partial destruction of the quasiparticle peaks.  

We thank K. Beach, N. Nagaosa, T.K Ng, T.M. Rice, M. Salmhofer, T. Senthil and A. Vishwanath for useful discussions. C.H. acknowledges fincancial support by Deutsche Forschungsgemeinschaft (DFG) and P.A.L. by NSF grant DMR-0201069.

\end{document}